\pdfoutput=1 
\documentclass[authoryear,10pt,5p,times]{elsarticle}

\usepackage{hyperref}

\journal{Astronomy and Computing}

\newcommand{\ac}{A\&C}
\newcommand{\q}[1]{`#1'}

\begin{document}
\begin{frontmatter}
\title{Astronomy and Computing: a new journal for the astronomical computing community\tnoteref{t1}}

\tnotetext[t1]{The authors of this paper are the editors of \ac.}

\author[aa]{Alberto Accomazzi}
\ead{aaccomazzi@cfa.harvard.edu}
\address[aa]{Harvard-Smithsonian Center for Astrophysics, 60 Garden Street, Cambridge, MA 02138, USA}

\author[tb]{Tam{\'a}s Budav{\'a}ri}
\ead{tamas.budavari@jhu.edu}
\address[tb]{Dept. of Physics \& Astronomy, The Johns Hopkins University, 3400 North Charles Street, Baltimore, MD 21218, USA}

\author[cf]{Christopher Fluke}
\ead{cfluke@astro.swin.edu.au}
\address[cf]{Centre for Astrophysics \& Supercomputing, Swinburne University of Technology, 1 Alfred Street, Hawthorn, 3122, Australia}

\author[ng]{Norman Gray}
\ead{norman@astro.gla.ac.uk}
\address[ng]{School of Physics \& Astronomy, University of Glasgow, Glasgow, G12 8QQ, UK}

\author[rgm]{Robert G Mann}
\ead{rgm@roe.ac.uk}
\address[rgm]{Institute for Astronomy, University of Edinburgh, Royal Observatory, Blackford Hill, Edinburgh, EH9 3HJ, UK}

\author[wo]{William O'Mullane}
\ead{womullan@sciops.esa.int}
\address[wo]{European Space Agency,
European Space Astronomy Centre,
E-28691 Villanueva de la Ca\~nada, Madrid, Spain}

\author[aw]{Andreas Wicenec}
\ead{andreas.wicenec@uwa.edu.au}
\address[aw]{International Centre for Radio Astronomy Research, University of Western Australia, 35 Stirling Highway, Crawley, WA 6009, Australia}

\author[mw]{Michael Wise}
\ead{wise@astron.nl}
\address[mw]{ASTRON (Netherlands Institute for Radio Astronomy), P.O. Box 2, 7990 AA Dwingeloo, The Netherlands}

\begin{abstract}
 We introduce \emph{Astronomy and Computing}, a new journal for
  the growing population of people working in the domain where astronomy overlaps with computer 
science and information technology. The journal aims to provide a new communication channel within
that community, which is not well served by current journals, and to help secure recognition of its true importance within modern astronomy. In this inaugural editorial, we describe the rationale for creating the journal, outline its scope and ambitions, and seek input from the community in defining in detail how the journal should work towards its high-level goals.

\end{abstract}

\end{frontmatter}

\section{Introduction}
\label{intro}

\emph{Astronomy and Computing} (\ac) is a new journal for the expanding community of people whose 
work focuses on the application of computer science and information technology within astronomy, rather than on astronomical research \emph{per se}. This domain is an increasingly important part of astronomy, but one that is poorly represented in the astronomical literature, resulting in inefficient sharing of knowledge within its community and a difficulty on the part of the members of that community to establish the track record of refereed publications needed for career advancement in many astronomical institutions. The over-riding goal of \ac\ is to address these two problems by providing a venue for the publication of peer-reviewed papers on astronomical computing that will act as a focus for the community, aiding its progress through the effective transmission of knowledge, and helping it secure recognition of its true value to astronomy. 

Some of the papers that will appear in \ac\ might otherwise have been
published in an existing journal, but most would not, for reason of
topic rather than quality. In Section~\ref{scope} we outline some of
the topics that fall within the scope of \ac, but that list is
illustrative, not exhaustive, and we have deliberately not attempted
to specify the boundaries of that scope too rigidly, because this is a
rapidly changing field.

The range of topics to appear will be mirrored by the range of people
writing and reading the papers. Some will be research astronomers with
content to share that is of a more technical nature than what is
usually published in the astronomical research literature. Some will be
academic computer scientists reporting on astronomy-centred projects
that are too \q{applied} to find a comfortable place in the computer science
literature. Many will have long track records in the application of
computational techniques and technologies to astronomy, whether they
started as astronomers, computer scientists or IT professionals. These
groups may disagree on what they call what they do~-- \q{astronomical
  software development}, \q{astroinformatics}, \q{astronomical
  computing}, \q{computational astronomy}, or more~--
but what will be common to the papers that appear in \ac\ is that they
focus on technical matters, not on presenting astronomical results.

In Section~\ref{ascom} we argue for the timeliness of the launch of \ac, whose aims and scope are outlined in Section~\ref{scope}. The topics listed there naturally lead to a set of different types of paper, as described in Section~\ref{types}. Some of these may have additional online content associated with them, which we discuss in Section~\ref{pragmatics}, along with other pragmatic issues relating to the publication of papers in \ac. Finally, Section~\ref{conclusions} rounds off this editorial with an invitation to the community to help make a reality of the vision outlined here.

\section{Astronomical computing as a discipline}
\label{ascom}

The proponents of any new journal must justify why it is needed, and why that need is currently so pressing as to actually lead to the launch of the journal. In this Section we make the case for the necessity of creating \ac, and the timeliness of doing so now, based on the maturity of its target domain and the critical mass of the community it will serve. 

\subsection{The justification for \ac}

Astronomy has long been a source of practically focused innovation.  With
the telescope, timekeeping, and computing, to name just three, astronomers have a proven record as early adopters of new technologies, often contributing generic innovation to these fields as well as acquiring the specific
skills necessary to exploit those technologies in support of their own
science.\footnote{The community has shown itself remarkably prescient.
In a mid-1980s \emph{Science} special issue \citep{brauman86} and in
\citep{wells87} the community identified key
technologies which have shown their longevity in practice.}  In the cases
of optics and instrumentation, and of precise timekeeping, as the technology
developed it became more specialized, more removed from the perceived mainstream
of astronomical research, and its specialists
stopped thinking of themselves as astronomers doing technology on the
side, and instead as technologists leading an independent sub-discipline
of the larger astronomical project.  That branching finds expression in
the largely separate, but still mutually intelligible, publishing world
of astronomical instrumentation, which has a thriving centre in the
SPIE conference series.

This branching off, into a technologically led sub discipline, has
already happened for astronomical computing, as reflected by the longevity of the
ADASS\footnote{\texttt{www.adass.org}} and ADA\footnote{\texttt{ada7.cosmostat.org}} conference 
series. One difference between astronomical instrumentation and
astronomical computing is that while the community undertaking the
former appears content with its current publication options, that
working in the latter domain is not, as evinced by the discussion at
the \q{Birds of a Feather} discussion at the 2010 ADASS conference in
Boston, summarized in \citep{gray11a}. 

This discussion effectively concluded that, since there were journals other
than the best-known ones, which professed willingness to accept
the submission of software-related papers, the community should and
would migrate to these.  This turned out to be overoptimistic, and
nothing like it appears to have happened in fact.
In any case, it is not clear
that colonizing an existing journal would give the discipline the
definition and visibility that it needs to help it grow, and we
believe that creating this disciplinary identity is as valuable a goal
as any search for professional credit.

\cite{gray11a} present three main reasons why the community can no
longer make do with a publication mechanism centred on the unrefereed
proceedings of an annual conference: (i)~a conference presents a
single submission deadline per year, forcing authors to publish when
the opportunity arises, not when the status of their project merits
it; (ii)~peer review can provide a quality threshold, and the
existence of guidelines will lead authors to justify and elaborate
their arguments to a greater degree, producing more comprehensive
papers; and (iii)~a journal~-- and especially a predominantly online
journal~-- will not have the space constraints that bedevil conference
proceedings, and so will allow authors to give their material the
detail it requires, and set it properly into its broad context of
previous work in a way that is impossible in a brief conference
report. However there is no intention that \ac\ will replace, say, the
ADASS proceedings volume: the two are complementary, serving different
community needs, and both outlets are necessary to ensure that those
needs are met in full.

For example, the institutions that employ many of \ac's intended
authors and readers find it harder to assess personal attainment on
the basis of a track record of successful projects than on a list of
refereed publications, so a peer-reviewed journal is needed to provide
vital support to career progression in this community.  That the
existing astronomical journals do not fulfil that role was
demonstrated by \citep{gray11a} whose authors circulated for comment
to the editors of the majority of them a set of abstracts of papers
from the previous year's ADASS conference (not all of those papers would
necessarily be appropriate for \ac~-- the point of the exercise was to
delineate for each journal the boundary in the computational domain
beyond which they would deem a paper to be too technical for their
audience).  The responses varied slightly between the main
astronomical journals, but the key finding from this exercise was
that, unsurprisingly, they view technical computational material as a
means to an end -- the justification of a scientific result -- rather
than an end in itself. Papers that focus on the technical material
will struggle to find a home in the existing astronomical journals,
and, when they do, they will be greatly outnumbered by \q{straight}
astronomical research papers, so these journals will not provide an
effective means of following the progress in astronomical
computing. This is an issue not only for the growing number of
astronomical computing specialists, but also for computer scientists
with expertise in areas (for example data mining or management of
\q{big data}) that overlap with astronomy and who need an interface to
the astronomical community. More importantly, this lack of a natural
home in the peer-reviewed literature leaves the technical material
underreported and underexplained: authors tone down the technical
detail and play up the associated astronomical result in order to get a
paper accepted, to the detriment of those wishing to understand and
build upon the technical lessons learnt.

\subsection{The timeliness of \ac}

There are several indicators that the astronomical computing community is reaching the critical mass that calls for the launch of a journal like \ac.  A number of new conference series  in this domain -- for example .Astronomy\footnote{\tt dotastronomy.com} and the international Astroinformatics conferences\footnote{See {\tt http://www.astro.caltech.edu/ai12/} for {\em Astroinformatics 2012}} -- are becoming established and accreting communities around themselves, while the {\em State of the Profession Position Papers\footnote{See {\tt http://sites.nationalacademies.org/BPA/BPA$\_$049492}}} submitted to the 2010 US Astronomy and Astrophysics Decadal Survey included a number of papers (such as~\cite{borne09}) highlighting its growing importance to astronomy. This has been also been recognized by the recent establishment by both the American Astronomical Society and the International Astronomical Union of working groups  covering the fields of astrostatistics and astroinformatics which are central to the scope of \ac.

This adoption of the word \q{astroinformatics}  reflects the use of analogous terms in other sciences: 
\emph{bioinformatics} has a secure existence at the analogous
intersection between biology and computing, absorbed by problems which
are too technical to be of much interest to most research biologists, and perhaps too
applied to be of interest to core computer science;
\emph{geoinformatics} has emerged in the same relation to geophysics.
It might be asked why it has taken so long for astroinformatics to 
emerge as a field, especially given the willingness of the astronomical community to innovate
in computational areas, as described earlier in this editorial. The
same question could be asked of experimental particle physics, where
the everyday work of many people who would still describe themselves
as particle physicists would be viewed by others as applied computing
in support of particle physics: they seem to view what they do as so
central to experimental particle physics that there is no need to
define an \q{X-informatics} subdiscipline. By analogy, some of the
authors of this editorial question the usefulness of the term
`astroinformatics' but, whatever the merits of the term, its growing
use is indicative of the increasing importance of computation to
astronomy, and underlines the timeliness of the launch of \ac.




\section{Aims and scope}
\label{scope}

Having argued that the time is right for a journal devoted to the
astronomical computing community, we now move on to discuss
what an \ac\ article should look like and what it should contain.

As the journal's web page\footnote{\url{http://www.journals.elsevier.com/astronomy-and-computing/}} says:
\begin{quotation}
\noindent \ac\ will focus on the broad area between
astronomy, computer science and information technology. The journal
aims to publish the work of scientists and (software) engineers in all
aspects of astronomical computing.
\end{quotation}

The web page also includes the following list of example topics:
\begin{itemize}

\item Scientific software engineering

\item Computational infrastructure

\item Computational techniques used for astrophysical simulations

\item Visualization

\item Data management, archives, and virtual observatory

\item Data analysis, data mining and statistics

 \item Data processing pipeline and automated systems

\item  Semantics, data citation and data preservation

\end{itemize}

This list is illustrative only:  we could easily
have listed double this number of topics, and we stopped only because a more
detailed list looks exhaustive, whereas a shorter one remains suggestive.
The defining feature of an \ac\ paper will be that it focuses on computation in support of astronomy, not on the astronomical results obtained using computation. 
We do not expect to find ourselves in much competition
with the existing set of core astronomy journals, nor with the more
theoretical areas of the computer science literature; instead we
are creating a complementary publication which will attract a class of
articles which has not hitherto had a clear identity or shape.

The detailed definition of this identity and shape will emerge in the first years of the
journal's existence, and we anticipate having more interplay
between authors, reviewers and editors than is usual in an
established journal, as we collectively work out the ideal structure
and content of an article in this area, and collectively identify what
is and what is not in scope for the field.

\section{Types of article}
\label{types}

Perhaps the best way of describing the journal's large scope is to
describe the range of articles that we anticipate seeing submitted,
and what we believe to be necessary or distinctive about them.  We
can identify at least the following broad categories, without necessarily being
committed to them indefinitely.


The most typical \textbf{research (or standard) articles} will describe an
innovative piece of work in the area, whether this
is a distinct project~-- a new algorithm, or system, or approach, or
application~-- or a major change in an established system, such as the
restructuring of an existing pipeline.  We expect to see a broad range
of articles in this category, but there are some particular species we
can identify from the outset.

One of these cases will be the \textbf{software release
articles}.   While a new major release of a piece of software, or a
library, will be a natural point at which to consider an \ac\ article,
it is not simply the increment of the version number that will
warrant publication, but perhaps the intellectual contributions of a
new algorithm, or the educative experience of a new software
engineering process, or novel technology.

While \ac\ will be a natural home for a \q{code
paper}, it will not act as a repository for code itself: we anticipate that a
software release that is
worth an \ac\ article will be one that is also worth being professionally packaged and
released at a stable URL, preferably with the source being additionally
available in a public code repository (and we suggest some suitable
repositories in the author instructions).  We regard such a distribution as
formally part of the article for the purposes of refereeing the submission.
Similarly, while we do not \emph{require} that a new algorithm be
accompanied by a released library implementation of it, we imagine that
a referee from the community would take some persuading that an algorithm
without such an implementation was nonetheless sufficiently well
corroborated by experience to warrant a scientific publication; while
we can imagine circumstances in which an algorithm would be described
without a \emph{public} implementation, we think it unlikely that it would be published
without at least the referee seeing the code.
Finally, while we as an editorial board have a
bias towards openly available code and open licences~-- if only
because they make things simpler 
for a community that may wish to incorporate, adapt and extend the
described code~-- we do not insist that code be licensed in such a
way.

Similarly, we believe that \ac\ will be a natural home for
\textbf{data release articles}, provided that these have significant technical content.  
The journal's scope gives the authors of such
articles the space to be as technically detailed as they
could want in their description of the development and delivery of a
new dataset, and might be natural counterparts to a simultaneous
astronomy article in another journal, which concentrates on the
science outputs.

We encourage the community to contribute \textbf{notes on practice}.  These
will be accounts of \q{lessons learned} in the course of trying and either
succeeding or failing with some technology or apparently promising
approach. These papers will be as formal, and as much of
a contribution, as a \q{standard} article, but with two important differences.
First, because they are notes on \emph{practice} we will expect
a different type of innovation from that necessary for a standard
research article.  That might be represented by the first,
or at least an early, application of a technology to a problem in
astronomy, or an application at a scale or in a fashion that represents
a significant commitment of intellectual energy.  Second, it should
be irrelevant whether the application of that technology succeeded or
failed: in either case, the project should be analysed in enough
detail, and at such a level of abstraction, that it would allow a reader to
understand \emph{why} the project succeeded or failed, and to be able
to use the information to predict with some confidence whether a
similar planned project would be likely to succeed
or fail in its turn. We expect those would most typically be technical
reasons, but the social, administrative or technical context is 
important as well: a particular innovation may fail simply because it was
applied in the wrong context, while it may succeed elsewhere.

Such a paper is arguably a type of \textbf{review article}.  As well as the
obvious analogues of review articles from other disciplines, we
imagine we will see similar, broadly pedagogical, accounts of
technologies applied to astronomical problems.  For example, we can
imagine that the early 2000s would have been a good time for a review
article about \q{Technological prospects and experience of XML in
Astronomy}, or the end of support for the IRAF or Starlink projects an
opportunity for an engineering review of project management lessons, or
we can imagine the~\citet{hleg10} producing a version of
their document as an \ac\ review article, or seeing a petascale
visualization review such as \cite{hassan11} here.

We will also consider \textbf{white papers}, describing some \q{state
  of the nation} in some respect, or plans for the future: for
example, some of the submissions for the recent US Decennial Review of
astronomy could be repurposed as \ac\ articles (for
example~\cite{borne09}), as might forward looks such
as~\cite{graham12} or~\cite{brescia12}.
While there may be
a number of such articles in the first few years of the journal's life, reflecting
the state of maturity of the domain it serves, these will be authoritative and
well grounded in expertise, and only accepted in response to an invitation from
the editorial board: the same goes for review articles, and we encourage authors
interested in writing one of either of these article types to contact an appropriate
editor to discuss the scope and approach of their proposed paper. All other types
of paper will be accepted as unsolicited contributions, although the editors are
always available to discuss possible papers with authors. 

As a form rather than a category, we will also introduce the
\textbf{target article}, which is familiar in some other academic
disciplines but not well known in the physical sciences. From time to time -- possibly
only once or twice a year -- the editors will identify that a submitted paper 
provides one side to an argument that is underway within the community, and will,
with the authors' permission, make the paper a target article. They will then solicit 
substantial commentaries on it (of perhaps one or two pages in length) and then 
publish in a single package the original article, its commentaries, and a final summary
from the target article's authors, thereby presenting a broader coverage of the topic
than would have resulted from the original paper alone.  We expect the
detailed form in this journal to evolve with experience.

Finally, we expect to publish occasional \textbf{special issues}, which may collect together
papers resulting from a specific conference, relating to a particular major project, marking some 
substantial milestone or event, or which, through some other connection, comprise a coherent whole 
that is greater than the sum of its parts.  
As above, we look forward to proposals from the community.

\section{Pragmatics}
\label{pragmatics}

This section discusses a number of more practical matters.

As noted above, \ac\ will not act as a code repository itself, nor will it
act as a data repository. However one of the key roles
of an academic journal is to act as the long-term archive of a
discipline's activities, and we are aware that this activity may be
expressed in the form of software which might not warrant a full-scale
distribution~-- illustrative codes, or perhaps templates~-- and
in other non-textual forms~-- we can imagine screencasts or
simulation fly-throughs, for example.  We have made no specific
provision for this at this stage in the journal's development, but
instead aim to develop the support for such artefacts, building on the
publisher's existing experience of this in other journals, when and as it
becomes clear what the community needs.  The potential support is
flexible, and so authors should feel free to offer supplementary
material accompanying their articles; again, editors will be glad to discuss
possibilities with authors. 

Any contemporary astronomy-related journal has to make
clear its relationship with arXiv and with the NASA Astrophysics Data System (ADS).  
\ac\ fully supports authors' use of arXiv: it supports authors submitting to the journal by 
providing the arXiv identifier of an already-uploaded preprint, and cooperates with
authors uploading post-refereeing preprints. \ac\ papers will be
indexed by ADS and will be findable through the traditional ADS metadata
search (`Abstract Search'). Additionally, the full content of the
papers will be indexed in ADS's new full-text search service, as part
of ADS's ongoing collaboration with Elsevier.  Thanks to the metadata
curation performed by ADS, the arXiv version of a paper (if one
exists) will be linked to its published record in ADS and, hence, be
accessible through it.

\ac\ will follow Elsevier's \q{article-based publishing} model, which
means that articles will be assigned final volume and page numbers,
will be assigned a DOI, and will appear in indexing services, as soon as they have
passed successfully through the peer-review process.

\section{Conclusions and an invitation}
\label{conclusions}

We believe that the launch of \emph{Astronomy and Computing}
represents an important stage in the emergence of \q{astronomical
computing} as a mature discipline within astronomy. We look
forward to the journal acting as a voice, arena and inspiration for
the community in the coming years and encourage the community
to join with us in making that a reality.

\def\eprint#1{\href{http://arxiv.org/abs/#1}{{\tt arXiv:#1}}}
\def\ArXivprefix{}
\bibliographystyle{model2-names}
\bibliography{ac-editorial}

\end{document}